\documentclass{article}
\usepackage[english]{babel}
\usepackage{spconf,amsmath,epsfig}
\usepackage{amsthm}
\usepackage{amsfonts}
\usepackage{amssymb}
\usepackage{graphicx,enumerate,multicol}
\usepackage[ruled]{algorithm2e}
\usepackage{aas_macros}

\newcommand{\eqi}{\gamma}

\newcommand{\kerm}{\mathrm{ker}}

\newcommand{\va}{\alpha}
\newcommand{\vx}{{x}}
\newcommand{\vy}{{y}}
\newcommand{\parenth}[1]{\left({#1}\right)}

\newcommand{\Pm}{\mathrm{P}}

\newcommand{\Id}{\boldsymbol{\mathrm{I}}}

\newcommand{\Hmb}{\boldsymbol{\mathrm{H}}}
\newcommand{\Kmb}{\boldsymbol{\mathrm{K}}}
\newcommand{\Lmb}{\boldsymbol{\mathrm{L}}}

\newcommand{\Phb}{\boldsymbol{\Phi}}

\newcommand{\opnorm}[1]{\left\lvert\!\left\lvert\!\left\lvert #1 \right\rvert\!\right\rvert\!\right\rvert}
\newcommand{\norm}[1]{\left\lVert #1 \right\rVert}

\newcommand{\abs}[1]{\left\lvert #1 \right\rvert}

\newcommand{\Cc}{\mathcal{C}}

\newcommand{\Hc}{\mathcal{H}}

\newcommand{\Kc}{\mathcal{K}}

\newcommand{\Mc}{\mathcal{M}}

\newcommand{\Pc}{\mathcal{P}}

\newcommand{\Prj}{\mathcal{P}}

\DeclareMathOperator{\prox}{prox}

\DeclareMathOperator{\dom}{dom}
\DeclareMathOperator{\ri}{ri}

\DeclareMathOperator*{\argmin}{argmin}

\renewcommand{\ge}{\geqslant}
\renewcommand{\le}{\leqslant}

\theoremstyle{plain}
\newtheorem{theorem}{\textbf{Theorem}} %[section]

\newtheorem{lemma}[theorem]{\textbf{Lemma}}

\newtheorem{propo}[theorem]{\textbf{Proposition}}

\newtheorem{definition}[theorem]{\textbf{Definition}}

\title{Inverse Problems with Poisson noise: Primal and Primal-Dual Splitting\footnote{Submitted to ICIP 2011 on the 01/21/11.}}
\name{F.-X. Dup\'e$^\text{a}$, M.J. Fadili$^\text{b}$ and J.-L. Starck$^\text{a}$}
\address{\small \begin{tabular}{cc}
    \begin{minipage}{0.5\linewidth}\begin{center} $^\text{a}$ AIM UMR CNRS - CEA \\ 91191 Gif-sur-Yvette France \end{center}\end{minipage} & 
    \begin{minipage}{0.5\linewidth}\begin{center} $^\text{b}$ GREYC CNRS-ENSICAEN-Universit\'e de Caen\\
       14050 Caen France\end{center} \end{minipage} \\
  \end{tabular}
}
\begin{document}
%\ninept
%
\maketitle
\begin{abstract}
%   Since the last ten years, many optimization methods have appears for most of the inverse problems. However, they are
%   generally different in the way they manage linear operators. For example, the Poisson deconvolution problem can be
%   solved using sparsity regularization using two different schemes: primal and primal-dual. Using this context, we
%   propose to compare these two schemes in term of convergence speed and times complexity and then propose some elements
%   of answer on which scheme is the best for the current problem.
In this paper, we propose two algorithms for solving linear inverse problems when the observations are corrupted by Poisson noise. A proper data fidelity term (log-likelihood) is introduced to reflect the Poisson statistics of the noise. On the other hand, as a prior, the images to restore are assumed to be positive and sparsely represented in a dictionary of waveforms. Piecing together the data fidelity and the prior terms, the solution to the inverse problem is cast as the minimization of a non-smooth convex functional. We establish the well-posedness of the optimization problem, characterize the corresponding minimizers, and solve it by means of primal and primal-dual proximal splitting algorithms originating from the field of non-smooth convex optimization theory. Experimental results on deconvolution and comparison to prior methods are also reported.
\end{abstract}
\begin{keywords}
 Inverse Problems, Poisson noise, Duality, Proximity operator, Sparsity.
\end{keywords}
\section{Introduction}
\label{sec:intro}

%Recently primal-dual methods have been successfully used for optimization tasks in inverse problems such as Poisson deconvolution \cite{Figueiredo2010} or TV-inpainting \cite{Chambolle2010}. These methods propose an elegant way for managing the linear operators involving the resolvent of the underlying functions and without any direct inversion of these operator. However, no study has been done on their difference with a primal method in terms of convergence speed and times complexity.

Linear inverse problems in presence of Poisson noise have attracted less interest in the literature than their Gaussian counterpart, presumably because the noise properties are more complicated to handle. Such inverse problems have however important applications in imaging such as restoration (e.g. deconvolution in medical and astronomical imaging), or reconstruction (e.g. computerized tomography). For instance, the well-known Richardson-Lucy has been proposed for deconvolution. The RL algorithm, however, amplifies noise after a few iterations, which can be avoided by introducing regularization. In \cite{Dey2004}, the authors presented a Total Variation (TV)-regularized RL algorithm, and \cite{Starck2006} advocated a wavelet-regularized RL algorithm.

In the context of Poisson linear inverse problems using sparsity-promoting regularization, a few recent algorithms have been
proposed. For example, \cite{Dupe2009c} stabilize the noise and proposed a family of nested schemes relying upon proximal splitting algorithms (Forward-Backward and Douglas-Rachford) to solve the corresponding optimization problem. The work of \cite{Chaux2009} is in the same vein. However, nested algorithms are time-consuming since they necessitate to sub-iterate. Using the augmented Lagrangian method with the alternating method of multipliers algorithm (ADMM), which is nothing but the Douglas-Rachford splitting applied to the Fenchel-Rockafellar dual problem, \cite{Figueiredo2010} presented a deconvolution algorithm with TV and sparsity regularization. This scheme however necessitates to solve a least-square problem which can be done explicitly only in some cases.   

In this paper, we propose a framework for solving linear inverse problems when the observations are corrupted by Poisson noise. In order to form the data fidelity term, we take the exact Poisson likelihood. As a prior, the images to restore are assumed to be positive and sparsely represented in a dictionary of atoms. The solution to the inverse problem is cast as the minimization of a non-smooth convex functional, for which we prove well-posedness of the optimization problem, characterize the corresponding minimizers, and solve them by means of primal and primal-dual proximal splitting algorithms originating from the realm of non-smooth convex optimization theory. Convergence of the algorithms is also shown. Experimental results and comparison to other algorithms on deconvolution are finally conducted.

%we propose two compares two optimization schemes which solve the same problem, here a Poisson deconvolution problem. One scheme is based on a primal algorithm which is a generalization of the Douglas-Rachford scheme to a sum of functions \cite{Combettes2008}. The other one, use the recent primal-dual approach proposed in \cite{Chambolle2010} for convex problems. We first stat the problem and then, describe both primal and primal-dual ways of solving the optimization problem. Finally, an experiment is performed in order to compare the behavior of the two deconvolution algorithms. \textbf{(Put some links on the results.)}

\subsection*{Notation and terminology}
\label{sec:notation}
Let $\Hc$ a real Hilbert space, here a finite dimensional vector subspace of $\mathbb{R}^n$. We denote by $\norm{.}$ the norm associated
with the inner product in $\Hc$, and $\Id$ is the identity operator on $\Hc$. $\norm{.}_p, p \geq 1$ is the $\ell_p$ norm. $\vx$ and $\va$ are
respectively reordered vectors of image samples and transform coefficients. We denote by $\ri \Cc$ the relative interior
of a convex set $\Cc$. A real-valued function $f$ is coercive, if $\lim_{\norm{\vx} \to +\infty}f\parenth{\vx}=+\infty$, and is proper if its domain is
non-empty $\dom f = \{ x\in\Hc \mid f(x) < +\infty \} \neq \emptyset$. $\Gamma_0(\Hc)$ is the class of
all proper lower semicontinuous (lsc) convex functions from $\Hc$ to $(-\infty,+\infty]$. We denote by $\opnorm{\mathbf{M}}= \max_{\vx \neq 0} \frac{\norm{\mathbf{M}\vx}}{\norm{\vx}}$ the spectral norm of the linear operator $\mathbf{M}$, and $\kerm(\mathbf{M}):=\{x \in \Hc: \mathbf{M}x=0, x \neq 0\}$ its kernel.
  
%\section{Sparse image representation}
%\label{sec:sparse-image-repr}

Let $x \in \Hc$ be an $\sqrt{n}\times\sqrt{n}$ image. $x$ can be written as the superposition of elementary atoms
$\varphi_\gamma$ parameterized by $\gamma \in \mathcal{I}$ such that $x = \sum_{\gamma \in \mathcal{I}} \alpha_\gamma
\varphi_\gamma = \Phb \va,\quad \abs{\mathcal{I}} = L, ~ L\ge n$. We denote by $\Phb: \Hc' \to \Hc$ the dictionary (typically a frame of $\Hc$), whose columns are the atoms all normalized to a unit $\ell_2$-norm
% In the rest of the paper, $\Phb$ will be an orthobasis or a tight frame with constant $A$.

\section{Problem statement}
\label{sec:problem-statement}

Consider the image formation model where an input image of $n$ pixels $\vx$ is indirectly observed through the action of a bounded linear operator $\Hmb: \Hc \to \Kc$, and contaminated by Poisson noise. The observed image is then a discrete collection of counts $\vy=(\vy[i])_{1 \le i
  \le n}$ which are bounded, i.e. $\vy \in \ell_{\infty}$. Each count $y[i]$ is a realization of an independent Poisson
random variable with a mean $(\Hmb \vx)_i$. Formally, this writes in a vector form as
{\small
\begin{equation}
  \label{eq:2}
  \vy \sim \Pc(\Hmb \vx)~.
\end{equation}
}
The linear inverse problem at hand is to reconstruct $\vx$ from the observed count image $\vy$.

A natural way to attack this problem would be to adopt a maximum a posteriori (MAP) bayesian framework with an
appropriate likelihood function {\textemdash} the distribution of the observed data $\vy$ given an original $\vx$
{\textemdash} reflecting the Poisson statistics of the noise. As a prior, the image is supposed to be economically
(sparsely) represented in a pre-chosen dictionary $\Phb$ as measured by a sparsity-promoting penalty $\Psi$ supposed
throughout to be convex but non-smooth, e.g. the $\ell_1$ norm.

From the probability density function of a Poisson random variable, the likelihood writes: 
{\footnotesize
\begin{equation}
  \label{eq:7}
  p(y|x) = \prod_i \frac{((\Hmb x)[i])^{y[i]} \exp\left(-(\Hmb x)[i]\right)}{y[i]!}~.
\end{equation}}
% and the associated log-likelihood function is
% \begin{equation}
%   \label{eq:8}
%   \ell\ell (y|x) = \sum_i \big( y[i]\log((\Hmb x)[i]) - (\Hmb x)[i] - \log(y[i]!)  \big)~.
% \end{equation}
% These formula are extended to the case $y=0$, using the convention $0!=1$. 
Taking the negative log-likelihood, we arrive at the following data fidelity term:
{\footnotesize
\begin{align}
  \label{eq:9}
  f_1\ &: \eta \in \mathbb{R}^n \mapsto \sum_{i=1}^n f_{\mathrm{poisson}}(\eta[i]), \\
  \text{if } y[i] > 0,\quad
  f_{\mathrm{poisson}}(\eta[i]) &=
  \begin{cases}
    -y[i] \log(\eta[i]) + \eta[i] & \text{if } \eta[i] > 0,\\
    +\infty & \text{otherwise,}
  \end{cases} \nonumber \\
  \text{if } y[i] = 0,\quad
  f_{\mathrm{poisson}}(\eta[i]) &=
  \begin{cases}
    \eta[i] & \text{if } \eta[i] \in [0,+\infty), \\
    +\infty & \text{otherwise.}
  \end{cases} \nonumber
\end{align}
}

Our aim is then to solve the following optimization problems, under a synthesis-type sparsity prior\footnote{Our framework and algorithms extend to an analysis-type prior just as well, though we omit this for obvious space limitation reasons.},
{\footnotesize
\begin{equation}
  \label{eq:11}
  \begin{gathered}
    \tag{$\Pm_{\eqi,\psi}$} \argmin_{\va\in\Hc'} J(\va), \\
    J\ :\ \va \mapsto {f_1\circ\Hmb\circ\Phb(\va)} + \eqi \Psi(\va) + \imath_{\Cc} \circ\Phb(\va)~.
  \end{gathered}
\end{equation}
}
The penalty function $\Psi : \va \mapsto \sum_{i=0}^{L} \psi_i(\va[i])$ is positive, additive, and chosen to enforce
sparsity, $\eqi > 0$ is a regularization parameter and $\imath_{\Cc}$ is the indicator function of the convex set
$\Cc$. In our case, $\Cc$ is the positive orthant since we are fitting Poisson intensities, which are positive by
nature.
%Notice, that an analysis-type prior may also be considered (see \cite{Figueiredo2010,Dupe2011} for more details).

From the objective in \eqref{eq:11}, we get the following,
{\small
\begin{propo}
\label{prop:objectives}
  {~}\\ \vspace*{-0.6cm}
  \begin{enumerate}[(i)]
    \setlength{\topsep}{0pt}
    \setlength{\parskip}{0pt}
    \setlength{\itemsep}{0pt}
    \setlength{\partopsep}{0pt}
  \item $f_1$ is a convex function and so are $f_1 \circ \Hmb$ and $f_1\circ\Hmb\circ\Phb$. 
  \item $f_1$ is strictly convex if $\forall i \in
    \{1,\ldots,n\}, y[i] \ne 0$. $f_1\circ\Hmb\circ\Phb$ remains strictly convex if $\Phb$ is an orthobasis and $\kerm(\Hmb) = \emptyset$.
  % \item The gradient of $f_1$ is
  %  \begin{align}
  %     \label{eq:104}
  %     \nabla f_1(\eta) &= (g(\eta[i]))_{1\le i\le n},\\
  %     \text{if } y[i] > 0,\quad
  %     g(\eta[i]) &=
  %     \begin{cases}
  %       1 - \frac{y[i]}{\eta[i]} & \text{if } \eta[i] > 0, \\
  %       +\infty & \text{else}.
  %     \end{cases} \nonumber \\
  %     \text{if } y[i] = 0,\quad
  %     g(\eta[i]) &=
  %     \begin{cases}
  %       1 & \text{if } \eta[i] \ge 0, \\
  %       +\infty & \text{else}.
  %     \end{cases} \nonumber  
  % \end{align}
    \item Suppose that $(0,+\infty) \cap \Hmb\left([0,+\infty)\right) \neq \emptyset$. Then $J \in \Gamma_0(\Hc)$.
  \end{enumerate}
\end{propo}
}

\subsection{Well-posedness of \eqref{eq:11}}
\label{sec:char-solut}

Let $\Mc$ be the set of minimizers of problem \eqref{eq:11}. Suppose that $\Psi$ is coercive. Thus $J$ is
coercive. Therefore, the following holds:
{\small
\begin{propo}
  {~} \\ \vspace{-0.5cm}
  \begin{enumerate}[(i)]
    \setlength{\topsep}{0pt}
    \setlength{\parskip}{0pt}
    \setlength{\itemsep}{0pt}
    \setlength{\partopsep}{0pt}
  \item Existence: \eqref{eq:11} has at least one solution, i.e. $\Mc\ne\emptyset$.
  \item Uniqueness: \eqref{eq:11} has a unique solution if $\Psi$ is strictly convex, or under (ii) of Proposition~\ref{prop:objectives}.
  \end{enumerate}
\end{propo}
}

\section{Iterative Minimization Algorithms}
\label{sec:sparse-iter-deconv}

\subsection{Proximal calculus}
We are now ready to describe the proximal splitting algorithms to solve \eqref{eq:11}. At the heart of the splitting framework is the notion of proximity operator.
{\small
\begin{definition}[\cite{Moreau1962}]
  \label{def:1}
  Let $F \in \Gamma_{0}(\Hc)$. Then, for every $x\in\Hc$, the function $y \mapsto F(y) + \norm{x-y}^{2}/2$
  achieves its infimum at a unique point denoted by $\prox_{F}x$. The operator $\prox_{F} : \Hc \to \Hc$
  thus defined is the \textit{proximity operator} of $F$.
\end{definition}
}
Then, the proximity operator of the indicator function of a convex set is merely its orthogonal projector. One important
property of this operator is the separability property:
{\small
\begin{lemma}[\cite{Combettes2005}]
  \label{lem:decomp}
  Let $F_k \in \Gamma_0(\Hc),\ k \in \{1,\cdots,K\}$ and let $G : (x_k)_{1\le k\le K} \mapsto \sum_k F_k(x_k)$. Then
  $\prox_{G} = (\prox_{F_k}) _{1 \le k \le K}$.
\end{lemma}
}

The following result can be proved easily by solving the proximal optimization problem in Definition~\ref{def:1} with
$f_1$ as defined in \eqref{eq:9}, see also \cite{Combettes2007a}.
{\small
\begin{lemma}
  \label{lem:prpois}
  Let $y$ be the count map (i.e. the observations), the proximity operator associated to $f_1$ (i.e. the Poisson anti
  log-likelihood) is,
    \begin{equation}
      \label{eq:3}
      \prox_{\beta f_1} \vx = \left( 
        \frac{\vx[i] - \beta + \sqrt{(\vx[i] -\beta)^2 + 4\beta \vy[i]}}{2}
      \right)_{1\le i \le n}~.
    \end{equation}
\end{lemma}
}

We now turn to $\prox_{\eqi\Psi}$ which is given by Lemma~\ref{lem:decomp} and the following result:
{\small
\begin{theorem}[\cite{Fadili2006}]
  \label{th:3}
  Suppose that $\forall~ i$: (i) $\psi_i$ is convex even-symmetric, non-negative and non-decreasing on $\mathbb{R}^+$,
  and $\psi_i(0)=0$; (ii) $\psi_i$ is twice differentiable on $\mathbb{R}\setminus \{0\}$; (iii) $\psi_i$ is continuous on $\mathbb{R}$, and admits a positive right derivative at zero ${\psi_i^{'}}_+(0) =
  \lim_{h\to 0^+} \frac{\psi_i(h)}{h} > 0$. Then, the proximity operator $\prox_{\delta\psi_i}(\beta) = \hat{\va}(\beta)$ has exactly one continuous solution decoupled in each coordinate $\beta[i]$ :
  \begin{equation}
    \label{eq:10}
    \hat{\va}[i] =
    \begin{cases}
      0 & \text{if } \abs{\beta[i]} \le \delta{\psi_i^{'}}_+(0)\\
      \beta_i-\delta\psi_i^{'}(\hat{\va}[i]) & \text{if } \abs{\beta[i]} > \delta{\psi_i^{'}}_+(0)
    \end{cases}
  \end{equation}
\end{theorem}
}
Among the most popular penalty functions $\psi_i$ satisfying the above requirements, we have $\psi_i(\va[i]) = \abs{\va[i]}, \forall ~ i$, in which case the associated proximity operator is soft-thresholding, denoted $\mathrm{ST}$ in the sequel.

\subsection{Splitting on the primal problem}
\label{sec:primal-method}

\subsubsection{Splitting for sums of convex functions}
Suppose that the objective to be minimized can be expressed as the sum of $K$ functions in $\Gamma_0(\Hc)$, verifying domain qualification conditions:
{\small
\begin{equation}
  \label{eq:sum}
  \argmin_{x \in \Hc} ~ \left(F(x) = \sum_{k=1}^K F_k(x)\right)~.
\end{equation}
}
Proximal splitting methods for solving \eqref{eq:sum} are iterative algorithms which may evaluate the individual proximity
operators $\prox_{F_k}$, supposed to have an explicit convenient structure, but never proximity operators of sums of the $F_k$.

Splitting algorithms have an extensive literature since the 1970's, where the case $K=2$ predominates. Usually, splitting algorithms handling $K > 2$ have either explicitly or implicitly relied on reduction of \eqref{eq:5} to the case $K = 2$ in the product space $\Hc^K$. For instance, applying the Douglas-Rachford splitting to the reduced form produces Spingarn's method, which performs independent proximal steps on each $F_k$, and then computes the next iterate by essentially averaging the individual proximity operators. The scheme described in \cite{Combettes2008} is very similar in spirit to Spingarn's method, with some refinements.

\begin{algorithm}[h]
  \small
  %\noindent{\bf{Task:}} Solve a linear inverse problem with Poisson noise. \\
  \noindent{\bf{Parameters:}} The observed image counts $y$, the dictionary $\Phb$, number of iterations
  $N_{\mathrm{iter}}$, $\mu > 0$ and regularization parameter $\eqi > 0$. \\
  \noindent{\bf{Initialization:}}\\
  $\forall i \in \{1,2,3\},\quad p_{(0,i)} = (0,0,0)^\mathrm{T}$. $z_0 = (0,0,0)^\mathrm{T}$. \\
  \noindent{\bf{Main iteration:}} \\
  \noindent{\bf{For}} $t=0$ {\bf{to}} $N_{\mathrm{iter}}-1$,
  \begin{itemize}
    \setlength{\topsep}{0pt}
    \setlength{\parskip}{0pt}
    \setlength{\itemsep}{0pt}
    \setlength{\partopsep}{0pt}
  \item \underline{Data fidelity} (Lemma~\ref{lem:prpois}): $\xi_{(t,1)}[1] = \prox_{\mu f_1/3}(p_{(t,1)}[1])$.
  \item \underline{Sparsity-penalty} (Lemma~\ref{th:3}): $\xi_{(t,1)}[2] =
    \prox_{\mu \eqi\Psi/3}(p_{(t,1)}[2])$.% = \mathrm{ST}_{\mu\eqi/3}(p_{(t,0)}[1])$.
  \item \underline{Positivity constraint}: $\xi_{(t,1)}[3] = \Prj_{{\Cc}}(p_{(t,1)}[3])$.
  \item \underline{Auxiliary constraints with $\Lmb_1$ and $\Lmb_2$:} (Lemma~\ref{th:prjli}):
    $\xi_{(t,2)} = \Prj_{{\ker \Lmb_1}}(p_{(t,2)}), \xi_{(t,3)} = \Prj_{{\ker \Lmb_2}}(p_{(t,3)})$.
  \item Average the proximity operators: $\xi_{t} = (\xi_{(t,1)} + \xi_{(t,2)} + \xi_{(t,3)})/3$. 
  \item Choose $\theta_t\in]0,2[$.
  \item Update the components: $\forall i \in \{1,2,3\},\quad p_{(t+1,i)} = p_{(t,i)} + \theta_t (2\xi_{t} - z_{t} - \xi_{(t,i)})$.
  \item Update the coefficients estimate: $z_{t+1} = z_t + \theta_t(\xi_t - z_t)$.
  \end{itemize}
  \noindent{\bf{End main iteration}} \\
  \noindent{\bf{Output:}} Reconstructed image $x^{\star}=z_{N_{\mathrm{iter}}}[0]$.
  \caption{Primal scheme for solving \eqref{eq:11}.}
  \label{algo:deconv}
\end{algorithm}

\subsubsection{Application to Poisson noise inverse problems}
Problem \eqref{eq:11} is amenable to the form \eqref{eq:sum}, by wisely introducing auxiliary variables. As \eqref{eq:11} involves two linear operators ($\Phb$ and $\Hmb$), we need two of them, that we define as $\vx_1 = \Phb\va$ and $\vx_2 = \Hmb\vx_1$. The idea is to get rid of the composition of $\Phb$ and $\Hmb$. Let the two linear operators $\Lmb_1 =[ \Id \quad 0 \quad -\Phb]$ and $\Lmb_2 = [ -\Hmb \quad \Id \quad 0]$. Then, the optimization problem \eqref{eq:11} can be equivalently written:
{\small
\begin{gather}
  \label{eq:1}
  \argmin_{(\vx_1,\vx_2,\va) \in \Hc\times\Kc\times\Hc'} \underbrace{f_1(\vx_2) + \imath_{\Cc}(\vx_1) + \eqi\Psi(\va)}_{G(\vx_1,\vx_2,\va)} + \\
  \imath_{\ker \Lmb_1}(\vx_1,\vx_2,\va) + \imath_{\ker \Lmb_2}(\vx_1,\vx_2,\va)~.
\end{gather}
}
Notice that in our case $K=3$ by virtue of separability of the proximity operator of $G$ in $x_1$, $x_2$ and $\alpha$; see Lemma~\ref{lem:decomp}.

The proximity operators of $F$ and $\Psi$ are easily accessible through Lemma~\ref{lem:prpois} and \ref{th:3}. The projector onto the positive orthant $\Cc$ is also trivial. It remains now to compute the projector on $\ker \Lmb_i$, $i=1,2$, which by well-known linear algebra arguments, is obtained from the projector onto the image of $\Lmb_i^*$.
{\small
\begin{lemma}
  \label{th:prjli}
  The proximity operator associated to $\imath_{\ker \Lmb_i}$ is
  \begin{gather}
    \label{eq:10a}
    \Prj_{\ker \Lmb_i} = \Id - \Lmb_i^* (\Lmb_i \circ \Lmb_i^*)^{-1} \Lmb_i~.
  \end{gather}
\end{lemma}
}
The inverse in the expression of $\Prj_{\ker \Lmb_1}$ is $(\Id + \Phb\circ\Phi^\mathrm{T})^{-1}$ can be computed efficiently when $\Phb$ is a tight frame. Similarly, for $\Lmb_2$, the inverse writes $(\Id + \Hmb\circ\Hmb^*)^{-1}$, and its computation can be done in the domain where $\Hmb$ is diagonal; e.g. Fourier for convolution. 

Finally, the main steps of our primal scheme are summarized in Algorithm~\ref{algo:deconv}. Its convergence is a corollary of \cite{Combettes2008}[Theorem~3.4].

{\small
\begin{propo}
Let $(z_t)_{t\in\mathbb{N}}$ be a sequence generated by Algorithm~\ref{algo:deconv}. Suppose that Proposition~\ref{prop:objectives}-(iii) is verified, and $\sum_{t\in\mathbb{N}} \theta_t(2-\theta_t) = +\infty$. 
%   \begin{enumerate}
%     \setlength{\topsep}{0pt}
%     \setlength{\parskip}{0pt}
%     \setlength{\itemsep}{0pt}
%     \setlength{\partopsep}{0pt}
%   \item (iii) of Proposition~\ref{prop:objectives}
%   \item $(0,\ldots,0)\in \mathrm{ri}\left\{ (z-z_1,\ldots,x-x_K)\mid z\in\Hc,\right.$\\
%       $\left.z_1\in\dom J_1,\ldots,z_K\in\dom J_K\right\}$;
%   \item $\sum_{t\in\mathbb{N}} \theta_t(2-\theta_t) = +\infty$;
%     % \item $\forall i \in  \{1,\ldots,K\}\quad \sum_{t\in\mathbb{N}} \theta_t \norm{a_{(t,i)}} < +\infty$.
%   \end{enumerate}
Then $(z_t)_{t\in\mathbb{N}}$ converges to a (non-strict) global minimizer of \eqref{eq:11}.
\end{propo}
}

\subsection{Splitting on the dual: Primal-dual algorithm}
\label{sec:primal-dual-method}
Our problem \eqref{eq:11} can also be rewritten in the form,
{\small
\begin{gather}
  \label{eq:5}
  \argmin_{\va \in \Hc'} F\circ\Kmb(\va) + \eqi\Psi(\va)
\end{gather}
}
where now $\Kmb = \begin{pmatrix} \Hmb\circ\Phb\quad \\ \Phb \end{pmatrix}$ and $F : (\vx_1,\vx_2) \mapsto f_1(x_1) + \imath_{\Cc}(x_2)$. Again, one may notice that the proximity operator of $F$ can be directly computed using the separability in $\vx_1$ and $\vx_2$.

Recently, a primal-dual scheme, which turns to be a pre-conditioned version of ADMM, to minimize objectives of the form \eqref{eq:5} was proposed in \cite{Chambolle2010}. Transposed to our setting, this scheme gives the steps summarized in Algorithm~\ref{algo:deconv2}.

Adapting the arguments of \cite{Chambolle2010}, convergence of the sequence $(\va_t)_{t\in\mathbb{N}}$ generated by Algorithm~\ref{algo:deconv2} is ensured.
{\small
\begin{propo}
  Suppose that Proposition~\ref{prop:objectives}-(iii) holds. Let $\zeta = \opnorm{\Phb}^2(1+\opnorm{\Hmb}^2)$, choose $\tau > 0$ and $\sigma$
  such that $\sigma\tau\zeta < 1$, and let $(\va_t)_{t\in\mathbb{R}}$ as defined by
  Algorithm~\ref{algo:deconv2}. Then, $(\va)_{t\in\mathbb{N}}$ converges to a (non-strict) global minimizer \eqref{eq:11} at the rate $O(1/t)$ on the restricted duality gap.
\end{propo}
}

\subsection{Discussion}
Algorithm~\ref{algo:deconv} and \ref{algo:deconv2} share some similarities, but exhibit also important differences. For
instance, the primal-dual algorithm enjoys a convergence rate that is not known for the primal algorithm. Furthermore,
the latter necessitates two operator inversions that can only be done efficiently for some $\Phb$ and $\Hmb$, while the
former involves only application of these linear operators and their adjoints. Consequently, Algorithm~\ref{algo:deconv2}
can virtually handle any inverse problem with a bounded linear $\Hmb$. In case where the inverses can be done
efficiently, e.g. deconvolution with a tight frame, both algorithms have comparable computational burden. In general, if
other regularizations/constraints are imposed on the solution, in the form of additional proper lsc convex terms that
would appear in \eqref{eq:11}, both algorithms still apply by introducing wisely chosen auxiliary variables.
%However, this would degrade the constant in the convergence speed of the primal-dual algorithm since the upper-bound in the proximal steps upper-bound would be smaller.

\begin{algorithm}[h]
  \small
  %\noindent{\bf{Task:}} Solve a linear inverse problem with Poisson noise. \\
  \noindent{\bf{Parameters:}} The observed image counts $y$, the dictionary $\Phb$, number of iterations
  $N_{\mathrm{iter}}$, proximal steps $\sigma > 0$ and $\tau > 0$, and regularization parameter $\eqi > 0$. \\
  \noindent{\bf{Initialization:}}\\
  $\va_0 = \bar{\va}_0 = 0$
  $\xi_0 = \eta_0 = 0$. \\
  \noindent{\bf{Main iteration:}} \\
  \noindent{\bf{For}} $t=0$ {\bf{to}} $N_{\mathrm{iter}}-1$,
  \begin{itemize}
    \setlength{\topsep}{0pt}
    \setlength{\parskip}{0pt}
    \setlength{\itemsep}{0pt}
    \setlength{\partopsep}{0pt}
  \item \underline{Data fidelity} (Lemma~\ref{lem:prpois}): $\xi_{t+1} = (\Id - \sigma \prox_{f_1/\sigma})(\xi_{t}/\sigma + \Hmb\circ\Phb \bar{\va}_{t})$.
  \item \underline{Positivity constraint}: $\eta_{t+1} = (\Id - \sigma\Prj_{{\Cc}})(\eta_{t}/\sigma + \Phb\bar{\va}_{t})$.
  \item \underline{Sparsity-penalty} (Lemma~\ref{th:3}):
    $\va_{t+1} = \prox_{\tau \eqi\Psi} \left(\va_t - \tau\Phb^\mathrm{T}\left(\Hmb^*\xi_{t+1} + \eta_{t+1}\right)\right)$.% = \mathrm{ST}_{\tau\eqi}(\va_t - \tau(\Phb^*\Hmb^*\xi_{t+1} + \Phb^*\eta_{t+1}))$
  \item Update the coefficients estimate: $\bar{\va}_{t+1} = 2\va_{t+1} - \va_t$
 \end{itemize}
  \noindent{\bf{End main iteration}} \\
  \noindent{\bf{Output:}} Reconstructed image $x^{\star}=\Phb\va_{N_{\mathrm{iter}}}$.
  \caption{Primal-dual scheme for solving \eqref{eq:11}.}
  \label{algo:deconv2}
\end{algorithm}

\section{Experimental results}
\label{sec:results}
\begin{figure}[htb]
  \centering
  \includegraphics[width=0.4\linewidth]{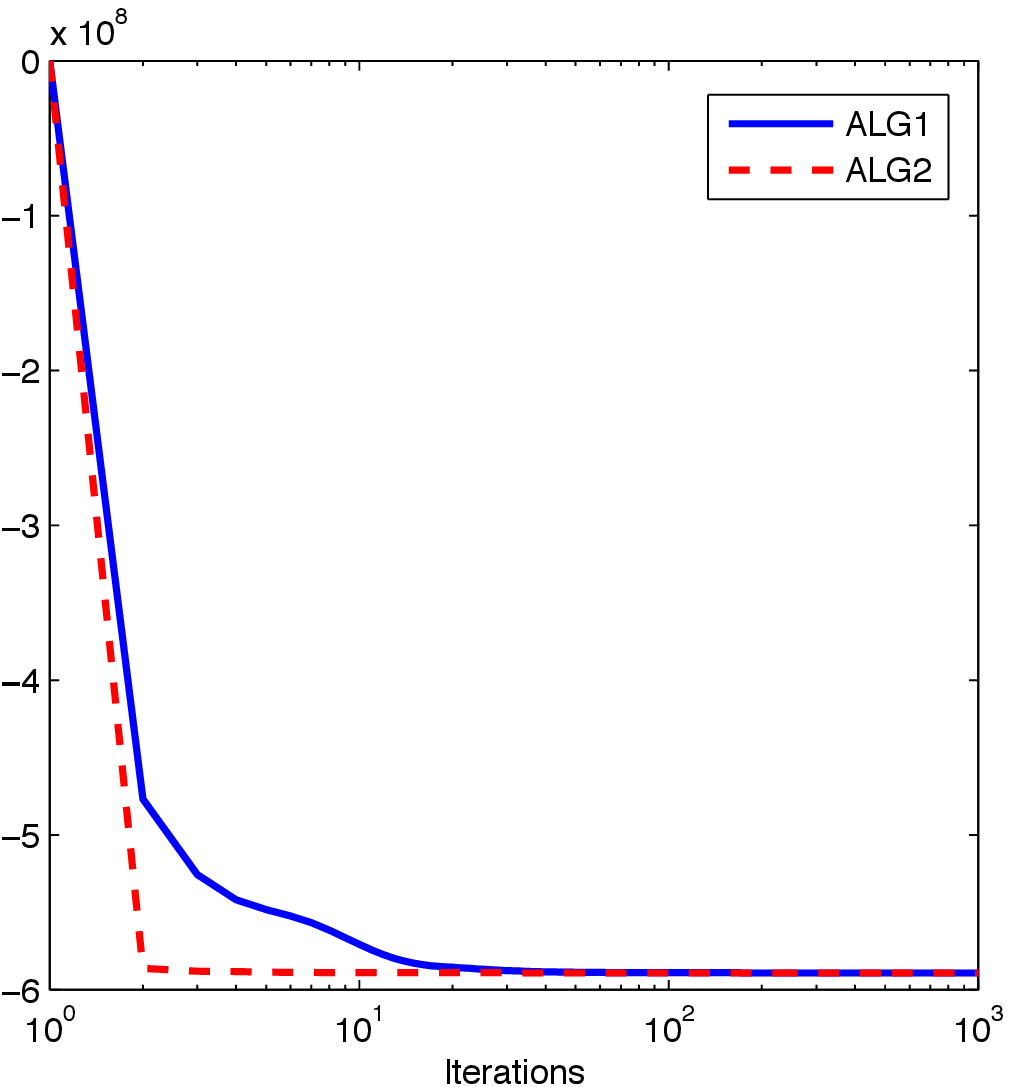}
  \includegraphics[width=0.4\linewidth]{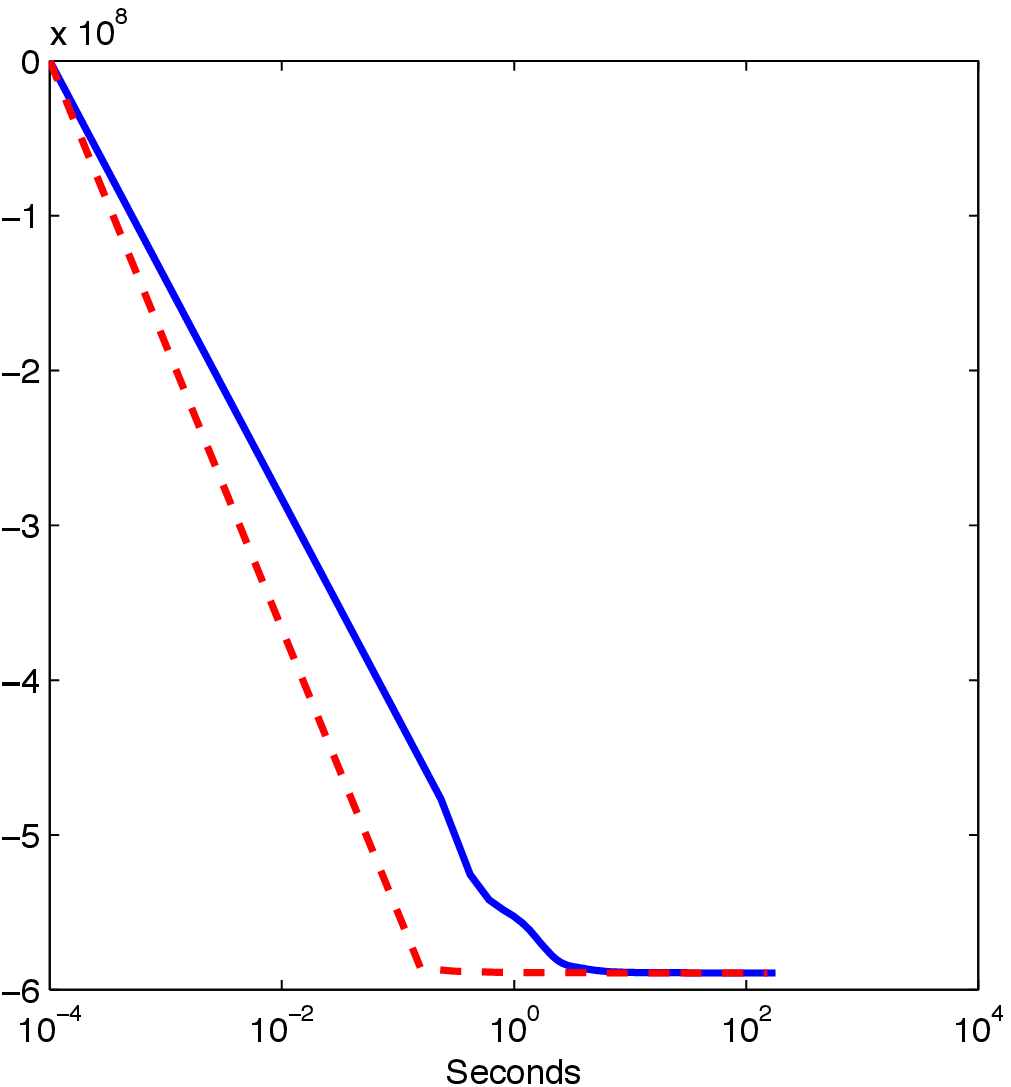}
  \caption{Objective function in function if iterations (left) and times (right).}
  \label{fig:plot}
\end{figure}

Our algorithms were applied to deconvolution. In all experiments, $\Psi$ was the $\ell_1$-norm. Table~\ref{tab:maesky}
summarizes the mean absolute error (MAE) and the execution times for an astronomical image, where the dictionary consisted of the wavelet transform and the PSF was that of the Hubble telescope. Our algorithms were compared to state-of-the-art alternatives in the literature. In summary, flexibility of our framework and the fact that Poisson noise was handled properly, demonstrate the capabilities of our approach, and allow our algorithms to compare very favorably with
other competitors. The computational burden of our approaches is also among the lowest, typically faster than the PIDAL algorithm. Fig.~\ref{fig:plot} displays the objective as a function of the
iteration number and time (in s). We can clearly see that Algorithm~2 converges faster than Algorithm~1.

%To assess the impact of the intensity level (average photon counts), we also report a comparative study in Table~\ref{tab:maeres}, where different algorithms were applied to an image of the planet Saturn at several intensity levels, with a $7 \times 7$ boxcar PSF. For this image, $\Phb$ was the curvelet dictionary. 

\begin{table}
\hspace*{-0.5cm}
  \centering
  {\footnotesize
    \begin{tabular}{|l||c|c|c|c|c|c|}\hline
      & RL-MRS \cite{Starck2006}& RL-TV \cite{Dey2004}& StabG \cite{Dupe2009c} & PIDAL-FS \cite{Figueiredo2010} & Alg.~\ref{algo:deconv} & Alg.~\ref{algo:deconv2} \\
      \hline MAE & 63.5      & 52.8    & 43  &  43.6 &  46     & 43.6  \\\hline
      Times & 230s & 4.3s & 311s & 342s & 183s & 154s \\ \hline
    \end{tabular}
  }
  \caption{\footnotesize MAE and execution times for the deconvolution of the sky image.}
  \label{tab:maesky}
\end{table}

% \begin{table}
%   \centering
%   {\footnotesize
%   \begin{tabular}{|c||c|c|c|c|c|c|}
%     \hline Intensity  Level & StabG & PIDAL-FS & Alg.~\ref{algo:deconv} & Alg.~\ref{algo:deconv2} \\
%     \hline 
%     \hline 5 	     & 1.14 (54.54)  & 0.24 (11.78)   & ()    & () \\
%     \hline 30        & 6.62 (52.78)  & 1.41 (11.23)   & ()    & () \\
%     \hline 100       & 26.73 (63.94  & 12.46 (29.81)  & ()    & () \\
% %    \hline 255       & 80.45 (75.46  & 49.36 (46.30)  & ()    & () \\
%     \hline
%   \end{tabular}
%   \caption{\footnotesize Average MAE values and relative MAE (in parentheses and in percent) for the Saturn image as a function of the intensity
%     level with the stabilization-based method (StabG)~\cite{Dupe2009c},
%     PIDAL-FS~\cite{Figueiredo2010}, and our two algorithms \ref{algo:deconv} and \ref{algo:deconv2}.}
%     }
%   \label{tab:maeres}
% \end{table}

\vspace*{-1.3cm}
\section{Conclusion}
\label{sec:conclusion}

In this paper, we proposed two provably convergent algorithms for solving the Poisson inverse problems with a
sparsity prior. The primal-dual proximal splitting algorithm seems to perform better in terms of convergence speed than the primal one. Moreover, its computational burden is lower than most comparable of state-of-art methods.

\footnotesize
\bibliographystyle{IEEEbib}
\bibliography{references}

\end{document}